\documentclass[prl,aps,twocolumn,groupedaddress,floatfix]{revtex4} 
\usepackage{concmath,charter}
\usepackage{amsmath} 
\usepackage{amsfonts}
\usepackage{amssymb}
\usepackage{bm}
\usepackage{verbatim} 
\usepackage[dvips]{graphicx} 
\usepackage[dvips]{color}
\usepackage{psfrag}
\usepackage{subfigure}

\newcommand{\e}[1]{\emph{#1}}

\newcommand{\avg}[1]{\langle {#1} \rangle}

\newcommand{\s}[0]{\sigma}

\newcommand{\Wh}[0]{\widehat {\mathcal W}}

\newcommand{\mW}[0]{\mathcal W}
\newcommand{\mP}[0]{\mathcal P}
\newcommand{\mQ}[0]{\mathcal Q}
\newcommand{\mPh}[0]{\widehat {\mathcal P}}
\newcommand{\mQh}[0]{\widehat {\mathcal Q}}

\begin{document}
\title{A simple stochastic model for the evolution of protein lengths}

\author{C. Destri$^1$ and C. Miccio$^1$}

\affiliation{ 1. Dipartimento di Fisica G.Occhialini, Universit\`a di
  Milano--Bicocca and INFN, Sezione di Milano, Piazza della Scienza 3
  - I-20126 Milano, Italy.}

\date{\today}

\begin{abstract}
  We analyse a simple discrete-time stochastic process for the theoretical
  modeling of the evolution of protein lengths. At every step of the
  process a new protein is produced as a modification of one of the
  proteins already existing and its length is assumed to be random variable
  which depends only on the length of the originating protein.  Thus a
  Random Recursive Trees (RRT) is produced over the natural integers. If
  (quasi) scale invariance is assumed, the length distribution in a single
  history tends to a lognormal form with a specific signature of the
  deviations from exact gaussianity. Comparison with the very large SIMAP
  protein database shows good agreement.
\end{abstract}

\maketitle

\section{Introduction}

Nowadays, it is well established that the great variety of proteins in
biological systems has been produced during the course of evolution by
means of gene mutations that take effect at the coding level
\cite{protevol1}. The main mechanisms are: duplication of genome segments
that contain sequences coding for one o more protein domains
\cite{dupl1,dupl2,dupl3,dupl4}; divergence of the duplicated sequences
by insertion, deletion and substitution of one or more base pairs
\cite{diverg1, diverg2, ins-del1, ins-del2}; domain rearrangements,
such as gene fusions and gene fission \cite{fus-fis1, fus-fis2},
domain recombination \cite{recomb1, recomb2}, gene shuffling
(recombination between dissimilar genes) \cite{geneshuff} and domain
insertions and deletions \cite{domins, domdel}. By means of these
microscopic mechanisms, iterated a huge number of times throughout the
ages of evolution, an initial protein population, most likely very
small and poorly assorted, has been enormously increased to the
present very large number and complex variety.

A valuable framework for the effective modeling of the evolutions of
genes and proteins could be provided by stochastic processes. In the
most general formulation, one should take into account the complex
organization of biological systems into independent organisms grouped
in turn into species, genera and kingdoms, as well as the complicated
effects of natural selection.  However, since all evolution mechanisms
generate new biological material by means of modifications of the
biological material already existing, in the case of proteins we may
imagine a simpler, more abstract discrete-time stochastic process over
the space of all amino acid sequences such that at each time step
$t=1,2,3,\ldots$ a new protein is generated with some prescribed
random mechanism from the set of proteins already existing.  Clearly
the dicrete time of the model has nothing to do with the time of the
true biological evolution process, except that it is a (almost)
monotonically increasing function of the latter, at least on time
scales large enough (great mass extinctions correspond most likely to
periods when this monotonicity is lost). Moreover, a single time step
in the process would correspond to some averaging over a multitude of
different effects, both at microscopic, or biochemical level, and at
the macroscopic level of the selection--based evolution mechanisms.

This abstract stochastic process would be specified by
$\mathrm{Pr}(p_{t+1}|\{p\}_t)$, that is the conditional probability
that the $t+1$ protein has the amino acid sequence $p_{t+1}$, given
that there is already a set $\{p\}_t\equiv\{p_1,p_2,\ldots p_t\}$ of
distinct proteins at time $t$. In principle,
$\mathrm{Pr}(p_{t+1}|\{p\}_t)$ might embody the effects of many, if
not all, of the complicated biochemical and evolutional mechanisms
alluded above and should depend explicitly on time. Moreover, the
initial configuration of proteins could be assumed to coincide with
the actual set of distinct amino acid sequences present in nature at
some moment in the distant past, when their total number was much
smaller than the present one. In any case, a huge amount of
information is required for a complete specification of the model
endowed with detailed predictive power.  On the other hand, all what
we may hope to reproduce in reasonably simple terms, to a large extent
independent on the details of the model, are some broad
characteristics of the distribution of protein currently observed.
This is indeed our main working hypothesis, based on the fact that the
very large number of proteins in the SIMAP database do show simple
universal properties \cite{simap,miccio}. Then we may rely on the basic
universality property, typical of a wide class of stochastic
processes, that is the capability to forget the details of the initial
transient regime and to relax toward a statistical equilibrium or
quasi-equilibrium state which depends only on very general features of
the conditional probability $\mathrm{Pr}(p_{t+1}|\{p\}_t)$ and is
characterized by few, weakly time--dependent ``macroscopic''
parameters.

\section{A stochastic process for protein lengths}\label{sec:stoch}

In the present work we concentrate our attention on the distribution of
protein lengths, that is the observed frequency of proteins with a specific
number of amino acids over the set of all known proteins. Thus we can
consider only the protein length as the random variable of the stochastic
process.  By definition this random length takes values in the natural
numbers and we denote it with the symbol $\ell$. We also observe that by
costruction all the proteins produced in the process can be ordered
according to the time of production, starting from $t=N_0$, with $N_0$ the
number of distinct proteins in the initial configuration, and arriving to
$t=N$, with $N$ of the order of the number of distinct proteins that exist
now in nature, which is of order $10^7$ or more. The statistical dynamics
of the process is fully determined by
$\mathrm{Pr}(\ell_{t+1}|\ell_{t},\ell_{t-1},\ldots,\ell_1)$, that is the
conditional probability that the $t+1$ protein has length $\ell_{t+1}$,
given that the preceeding proteins have the indicated lengths. This
conditional probability could depend explicitly on the formal time $t$.

As already discussed above in a more general context, the detailed
biological mechanisms that constrain
$\mathrm{Pr}(\ell_{t+1}|\ell_{t},\ell_{t-1},\ldots,\ell_1)$ are far
too complex to be explicitly incorporated in the model. Therefore, we
shall make simple and workable assumptions on the conditional
probability, relying, in practice, on some sort of central limit
theorem for the probability that a protein taken at random from the
state of a very long random process has a certain length.

 As a first simplifying assumption on the conditional probability
$\mathrm{Pr}(\ell_{t+1}|\ell_{t},\ell_{t-1},\ldots,\ell_0)$ we make that of
{\em locality}. That is, we assume that a given
protein length can be produced from a preceding length independently on
all the other lengths already produced. Hence we can write
\begin{equation}\label{eq:locality}
  \mathrm{Pr}(\ell_{t+1}|\ell_{t},\ell_{t-1},\ldots,\ell_1) = 
  \sum_{s=1}^t q_s \, W_s(\ell_{t+1}|\ell_{s})
\end{equation}
where the nonnegative weights $q_s$ are properly normalized,
$\sum_{s=1}^t q_s = 1$, and $W_s(\ell|\ell')$ can be interpreted as
Markovian transition probabilities.  In the absence of any other
information, one would assume equal a priori probabilities among the
different proteins, that is $q_s=1/t$ and
$W_s(\ell|\ell')=W(\ell|\ell')$ with no explicit $s-$dependence. This
might appear in conflict, however, with the global changes of
ecosystems as well as with the complex organization of biological
systems in kingdoms and species (which suggests that all proteins
existing nowadays can be roughtly divided into subsets of similar
proteins having almost independent evolutional histories, as least not
too far in the past). We may take this into account by restricting the
predictions of our stochastic process to suitable chosen subsets of
the proteins of the SIMAP database, according, for instance, to given
kingdoms. Moreover, we can neglect, on average, the global changes of
ecosystems by placing the start of the process not too deep in the
past. Alltogether let us assume that
\begin{equation}\label{eq:loc-hom}
  \mathrm{Pr}(\ell_{t+1}|\ell_{t},\ell_{t-1},\ldots,\ell_1) = 
  \frac1{t}\sum_{s=1}^t W(\ell_{t+1}|\ell_{s})
\end{equation}
Our stochastic process now differs from a random walk on the natural
integers only because at each step anyone of the already existing lengths,
rather than only the last generated one, may serve as starting point for a
jump to a new length. We are therefore dealing with the so--called Random
Recursive Tree (RRT) \cite{moon,burda} (more precisely, a random recursive
forest) embedded by $W(\ell|\ell')$ into the natural integers. It follows
that the probability $P(\ell,t)$ that the $t-$th protein has exactly length
$\ell$ satisfies the non--Markovian evolution equation
\begin{equation}\label{eq:evolution}
  P(\ell,t+1) = \sum_{\ell'=1}^\infty W(\ell|\ell')\,Q(\ell',t)
\end{equation}
where $Q(\ell,t)$ is the average length distribution (that is the mean
fraction of proteins of length $\ell$) at time $t$ and therefore evolves as
\begin{equation}\label{eq:Nevolution}
  (t+1)\,Q(\ell,t+1) = t\,Q(\ell,t) + P(\ell,t+1)
\end{equation}
Together eqs.~(\ref{eq:evolution}) and~(\ref{eq:Nevolution}) define a
stochastic process with memory and should be compared to the Markov chain
recursion for a simple Random Walk (RW) on all possible lengths, which
would read instead
\begin{equation}\label{eq:markov}
  P^{\rm RW}(\ell,t+1) = 
  \sum_{\ell'=1}^\infty W(\ell|\ell')\,  P^{\rm RW}(\ell',t)
\end{equation}
without any memory of the past. We still do have to make some choice
on the explicit form of $W(\ell|\ell')$, in which case our stochastic
process could be quite easily simulated on a computer. We expect,
however, that the large time asymptotic regime of the process depends
only on very general features of functional form of $W(\ell|\ell')$,
again thanks to the universality hypothesis which has its roots in the
law of large numbers. In any case, to investigate the statistics of
the produced lengths, we need beforehand some useful
properties and formal manipulations valid for any $W(\ell|\ell')$.

We recall first of all that by definition the nonnegative numbers
$W(\ell|\ell')$ satisfy the normalization condition:
\begin{equation*}
  \sum_{\ell=1}^\infty W(\ell|\ell') = 1
\end{equation*}
These transition probabilities are the elements of a matrix $W$, the
so--called stochastic matrix in the case of Markov processes.  Without loss
of generality, we may take $W$ to be ergodic, that is such that any finite
length can be produced after a suitable number of steps starting from any
other finite length.

Next we can exploit the linearity of eq.~(\ref{eq:evolution}) to
simplify the choice of initial conditions for $P(\ell,t)$. As already
stated above, the process is assumed to start with $N_0$ distinct
proteins, which we may take to have $n$ distinct lengths $\ell_j$,
$j=1,2,\ldots,n$, each repeated $n_j$ times so that $\sum_jn_j=N_0$.
This defines the initial length distribution
\begin{equation*}
  Q(\ell,N_0) = \frac1{N_0}\sum_{j=1}^n  n_j\,\delta_{\ell\,\ell_j}
\end{equation*}
when $N_0$ was the total number of distinct proteins. As the process may
start from anyone of these initial proteins with equal probability
$1/N_0$, we may regard $n_j/N_0$ as probability that the process starts
exactly from the length $\ell_j$. Therefore the solution of
eq.~(\ref{eq:evolution}) can be written
\begin{equation}\label{eq:superpose}
  \begin{split}
    P(\ell,t) &= \frac1{N_0}\sum_{j=1}^n n_j \, P(\ell,t-N_0+1|\ell_j)\\
    &= \sum_{\ell'=1}^\infty P(\ell,t-N_0+1|\ell') \,  Q(\ell',N_0)
  \end{split}
\end{equation}
where $P(\ell,t|\ell')$ is the special solution that is concentrated on the
arbitrary value $\ell'$ at $t=1$, that is
$P(\ell,1|\ell')=\delta_{\ell\,\ell'}$. Similarly we have 
\begin{equation}\label{eq:Qsuperpose}
  Q(\ell,t) = \sum_{\ell'=1}^\infty Q(\ell,t-N_0+1|\ell')\,Q(\ell',N_0)
\end{equation}
where $Q(\ell,t|\ell')$ is the solution of eq.~(\ref{eq:Nevolution})
specialized to $P(\ell,t|\ell')$, that is
\begin{equation}\label{eq:nuvsP}
  Q(\ell,t|\ell') = \frac1{t}\sum_{s=1}^{t} P(\ell,s|\ell')
\end{equation}
Clearly $\ell'$ is the length of a specific protein which plays the role of
root for the RRT, while the complete process is a forest of RRT's each
having root in one of the $N_0$ initial proteins and growing in
parallel. That is, the unique protein labelled by $t$ has a fixed
probability  $n_j/N_0$ of belonging to the tree rooted in a protein of length
$\ell_j$. 

We may now introduce the matrix notation
\begin{align*}
  W(\ell|\ell') & \equiv \big[W\big]_{\ell\,\ell'}  \\
  P(\ell,t|\ell') & \equiv \big[P(t)\big]_{\ell\,\ell'}  \\  
  Q(\ell,t|\ell') & \equiv \big[Q(t)\big]_{\ell\,\ell'}
\end{align*}
which allows to write the evolution equation for $P(t)$ more compactly as
\begin{equation}\label{eq:matevol}
  P(t+1) = W Q(t) = \frac1{t}\sum_{s=1}^{t}W P(s)
\end{equation}
Or equivalently as 
\begin{equation}\label{eq:matevol2}
  t\,P(t+1) = \sum_{s=1}^{t-1}W P(s) + W P(t) = (t+W-1)P(t)
\end{equation}
which has the formal solution
\begin{equation}\label{eq:tformal}
  P(t) = \prod_{s=1}^{t-1} \left(1 + \frac{W-1}{s} \right) =
  \frac{W^{\overline{t-1}}}{(t-1)!}
\end{equation}
where $z^{\bar n}$ stands for the so-called {\em raising factorial product}
$z(z+1)\ldots(z+n-1)$~\cite{concretem}. The raising factorial generates the
(unsigned) Stirling numbers of the first kind as coefficients of its
expansion in simple powers of $z$:
\begin{equation*}
  z^{\bar n} = \sum_{k=1}^n\Big[\begin{matrix}n\\k\end{matrix}\Big]
  \, z^k \;, \quad n>0
\end{equation*}
where we adopted the square bracket notation of ref.\cite{concretem} for
the Stirling numbers. Hence from eq.~(\ref{eq:tformal}) we can write
\begin{equation}\label{eq:Pstirling}
   P(t) = \frac1{(t-1)!}
   \sum_{s=1}^{t-1}\Big[\begin{matrix}t-1\\s\end{matrix}\Big] P^{\rm RW}(s)
\end{equation}
where $P^{\rm RW}(t)=W^t$ is the formal solution of the standard Random
Walk. Notice that, by eq.~(\ref{eq:Pstirling}), $P(t)$ is indeed properly
normalized, that is
\begin{equation*}
  \sum_{\ell=1}^{\infty} P(\ell,t|\ell') = 1
\end{equation*}
since $W^s$ is also a stochastic matrix and  Stirling numbers satisfy the
normalization 
\begin{equation*}
  \sum_{k=1}^n\Big[\begin{matrix}n\\k\end{matrix}\Big] = n! \;, \quad n>0
\end{equation*}
In fact, eq.~(\ref{eq:Pstirling}) shows that the quantity
\begin{equation*}
  p_s(t) = \frac1{(t-1)!}\Big[\begin{matrix}t-1\\s\end{matrix}\Big] 
\end{equation*}
has the interpretation, for the abstract RRT, of probability that the node
added at time $t$ is at a chemical distance $s$ from the root of the tree,
that is the original node present at $t=1$. In terms of proteins,
$p_s(t-N_0+1)n_j/N_0$ is therefore the probability that the $t-$th protein is
obtained through $s$ changes from one of the $N_0$ initial proteins of length
$\ell_j$.

Notice also that the evolution equation (\ref{eq:matevol}) allow to
write an alternative expression for $Q(t)$ which is local in time (but
generally non local in ``space'') w.r.t. $P(t)$
\begin{equation}\label{eq:P2Q}
   Q(t) = W^{-1}P(t+1) =  \frac1{t!}\sum_{s=1}^t
   \Big[\begin{matrix}t\\s\end{matrix}\Big] \,W^{s-1}
\end{equation}
One can see that $p_{s+1}(t+1)$, which by construction satisfies
\begin{equation*}
  p_{s+1}(t+1) = \frac1{t}\sum_{k=1}^t p_s(k)
\end{equation*}
represents the average fraction of nodes at a distance $s$ from the
root\cite{moon}.

For very large $n$ we can use the approximation
\begin{equation}\label{eq:euler}
   z^{\bar n} \simeq
  \frac{\Gamma(n)}{\Gamma(z)} \,n^z\Big[1+O\Big(\frac1{n}\Big)\Big]
\end{equation}
which follows from Euler's infinite product representation of the $\Gamma$
function \cite{abrste}. From eqs.~(\ref{eq:tformal}),~(\ref{eq:P2Q})
and~(\ref{eq:euler}) we then find
\begin{equation}\label{eq:nularget}
   Q(t) \simeq \frac{\exp[(W-1)\log t]} {\Gamma(W+1)}
\end{equation}
where we neglected all inverse powers of $t$ in the exponent, relying for
uniformity on the boundedness of its spectrum of $W$. The crucial point of
eq.~(\ref{eq:nularget}) is the very slow logarithmic dependence on time,
which appear evident upon comparison with the formal solution
$W^t=\exp(t\log W)$ of the Markovian case. 

In order to provide more explicit expressions for $P(t)$ and $Q(t)$ we
need some special assumtion on the stochastic matrix $W$. We do that in the
next section.

\section{Average properties of scale--invariant models} 

We describe here a class of examples which can be treated in detail at the
analytic level. These are characterized by the assumption that our
stochastic process is (almost) scale invariant. Intuitively, one expects
that longer proteins can be changed throughout evolutions more easily than
shorter ones. Exact scale invariance would mean that changes are
proportional to length.

To implement this picture, we first extend the lengths $\ell$ from the
positive integers to all real positive numbers. It will become
apparent in the sequel that this extension has very little impact on our
conclusions.  Next we change variables, from $\ell$ to its logarithm
$x=\log\ell$ and assume homogeneity in $x$, namely that
\begin{equation*}
  W(\ell|\ell')\,d\ell = W(e^x|e^{x'}) \, d(e^x) = \mW(x-x')\, dx 
\end{equation*}
is translation invariant, {\em i.e.} function only of $x-$space
differences. The process is very simple: at each time step the random
walker may pick anyone of the previously visited points as starting point
for the next step, whose value $x$ is extracted with the one--step pdf
(probability distribution function) $\mW(x)$. In terms of protein lengths,
at each step the length is rescaled by a factor $e^x$.  Since the true
variables are discrete, we may take $\mW(x)$ to be very smooth for all $x$.
Likewise, since $\ell\ge1$ we may take $\mW(x)$ to vanish very quickly (let
us say ``faster than any power'') for $x\to -\infty$. For $x\to +\infty$ we
assume instead quite reasonably that $\mW(x)$ vanishes fast enough to have
finite moments at least up to order four. We then introduce the follwing
notation for the first two cumulants:
\begin{equation*}
  \mu = \int dx \,x\,\mW(x) \;,\quad \sigma^2 = \int dx \,(x-\mu)^2\,\mW(x)
\end{equation*}
that is the mean value and the squared fluctuations.

We can now define the process probability in $x-$space as
\begin{equation*}
  \mP(x-x',t)\equiv e^x\,P(e^x,t|e^{x'})
\end{equation*}
and in the same way we can introduce the average distribution $\mQ(x,t)$
which by eq.~(\ref{eq:nuvsP}) satisfy
\begin{equation*}
  \mQ(x,t) = \frac1{t}\sum_{s=1}^{t} \mP(x,s)
\end{equation*} 
Since the stochastic matrix which correspond to $\mW(x-x')$ is diagonal in
Fourier space, we can now write the formal expression
eq.~(\ref{eq:tformal}) as
\begin{equation}\label{eq:ftrans}
  \mP(x,t) = \int\frac{dk}{2\pi} \,e^{i k x} \mPh(k,t)
\end{equation}
where
\begin{equation*}
   \mPh(k,t) = \prod_{s=1}^{t-1} \left[1 + \frac{\Wh(k)-1}{s} \right]
\end{equation*} 
and $\Wh(k)$ is the Fourier transform of $\mW(x)$.  Clearly, by
eq.~(\ref{eq:P2Q}), the Fourier transform of $\mQ(x,t)$ reads
\begin{equation*} 
  \mQh(k,t) = \frac{\mPh(k,t+1)}{\Wh(k)}
\end{equation*}
The correct normalization of either $\mP(x,t)$ or $\mQ(x,t)$ follows from
that of $W(x)$, which implies $\Wh(0)=1$. Other consequences of the
probabilistic nature of $W(x)$ is the symmetry $\Wh(k)^\ast=\Wh(-k)$ and
the bound $|\Wh(k)|\le 1$. In addition, with the natural requirements made
above on the one--step pdf $\mW(x)$, the function $\Wh(k)$ has the
expansion near $k=0$
\begin{equation}\label{eq:kdevel}
  \Wh(k) \simeq 1 -i\mu k - \tfrac12(\mu^2+\sigma^2)k^2 + \ldots
\end{equation}
and vanishes for large $|k|$.

In this context of a continuous logspace, the extension to a generic initial
distribution is very simple: it amounts to multiply both Fourier 
transforms $\mPh(k,t)$ and $\mQh(k,t)$ by the Fourier transform of the
initial distribution.

Using eq.~(\ref{eq:nularget}) we have for large $t$
\begin{equation}\label{eq:finteg}
  \mP(x,t) = \int\frac{dk}{2\pi} \,e^{ikx} \,\frac{
      \exp[(\Wh(k)-1)\log t]} {\Gamma(\Wh(k))}
\end{equation} and 
\begin{equation}\label{eq:nuinteg}
  \mQ(x,t) = \int\frac{dk}{2\pi} \,e^{ikx} \,\frac{
      \exp[(\Wh(k)-1)\log t]} {\Gamma(\Wh(k)+1)}
\end{equation}
up to fully negligible inverse power corrections in $t$. For any given
$\Wh(k)$ the Fourier integral in eqs.~(\ref{eq:finteg})
and~(\ref{eq:nuinteg}) can be computed numerically to high accuracy through
Fast Fourier Transform. Moreover, for large $t$ we can derive very similar
asymptotic expansions in inverse powers of $\log t$ valid for any $\Wh(k)$
in the class described above. Since our main interest is in the average
distribution profile $\mQ(x,t)$ we concentrate on this our attention.

The leading term for large $t$ is determined only by the first two terms of
the $\Wh(k)$ expansion (\ref{eq:kdevel}) near $k=0$, with the quadratic
term providing the Gaussian dominance in eqs.~(\ref{eq:finteg})
and~(\ref{eq:nuinteg}) accordig to the central limit theorem. From the
first and second derivatives in $k=0$ of Fourier transform $\mQh(k,t)$, we
first compute the mean value and standard deviation of the process for
large $t$, as
\begin{equation}\label{eq:avgdelta}
  \begin{split}  
    {\bar\mu}_t &\equiv \avg{x}_t = \mu(\log t + \gamma -1) \\
    {\bar\sigma}_t^2 &\equiv \avg{(x-\mu_t)^2}_t\\ &= 
    (\mu^2+\sigma^2)(\log t + \gamma-1) -\big(\tfrac{\pi^2}{6}-1\big)\mu^2
  \end{split}
\end{equation}
where $\gamma=0.5772156\ldots$ is Euler--Mascheroni constant.
Then in terms of the standard centerd scaled variable
\begin{equation*}
  \xi = \xi(x,t) = \frac{x-{\bar\mu}_t}{{\bar\sigma}_t}
\end{equation*}
we have to leading order
\begin{equation}\label{eq:leading}
  \mQ(x,t) \simeq \frac{e^{-\xi^2/2}}{\sqrt{2\pi {\bar\sigma}_t^2}}
\end{equation}
Going back the length variable $\ell$ through the definition
$x=\log(\ell/\ell')$, we find the lognormal distribution
\begin{equation}\label{eq:Pleading}
  Q(\ell,t|\ell') \simeq \frac{e^{-[\log\ell-\log(\ell'e^{{\bar\mu}_t})]^2/
      (2\bar\sigma_t^2)}}{\ell\,\sqrt{2\pi {\bar\sigma}_t^2}}
\end{equation}   
peaked around the rescaled initial length $\ell'\,e^{\bar\mu_t-\bar\sigma^2_t}$.
  
Subleading contributions to the above results, of relative order
$1/\sqrt{\log t}$ and smaller, {\em at fixed} values of $\xi$, can be
computed by the standard perturbation technique around Gaussian integrals:
one includes also terms of order higher than $k^2$, say of order $n\ge4$,
in the power expansion of $\Wh(k)$ around $k=0$ and then expands to order
$k^n$ also the exponentials of such terms; for completeness, also the
expansion of the inverse $\Gamma$ function must be properly extended;
finally one integrates explicitly each term of the complete expansion in
terms of multiple derivatives of the leading Gaussian. One obtains in this
way a $n-$degree polynomial in $\xi$ times the Gaussian $e^{-\xi^2/2}$. The
$n+1$ coefficients of the polynomial are fixed by the first $n+1$ moments
of the distribution, which in turn can be computed directly from the Taylor
series in $k=0$ of the Fourier transforms $\mQh(k,t)$ or $\mPh(k,t)$ (by
construction, we must impose $\avg{\xi}_t=0$ and $\avg{\xi^2}_t=1$ for the
first two moments). Taking into account the specific form of these Fourier
transforms, it is more convenient to calculate the cumulants of $\mQ(x,t)$
or $\mP(x,t)$, since their $n-$order cumulant is given by $\log t$ times
the $n-$order {\em moment} of the one--step pdf $\mW(x)$, plus the
$n-$order derivative w.r.t. $k$ of $\log[\Gamma(\Wh(k)+1)]$ or
$\log[\Gamma(\Wh(k))]$ evaluated at $k=0$. Moreover, the latter
contributions are systematically subleading as compared to the moments of
$\mW(x)$, so that we have, for the third--order and the fourth--order
cumulant of $\mQ(x,t)$ (that is the average skewness ${\bar s}_t$ and
kurtosis ${\bar\kappa}_t$ of the process, up to normalization conventions):
\begin{equation}\label{eq:hcumul}
  \begin{split}
   & {\bar s}_t \equiv \avg{\xi^3}_t = \frac{\mu_3}{\mu_2^{3/2}} 
   \frac1{\sqrt{\log t}} 
   \,\left[ 1+ {\cal O}\Big(\frac1{\log t}\Big) \right] \\
   & {\bar\kappa}_t \equiv\avg{\xi^4}_t - 3 = \frac{\mu_4}{\mu_2^2} 
   \frac1{\log t} \,\left[ 1+ {\cal O}\Big(\frac1{\log t}\Big) \right]
  \end{split}
\end{equation} 
where $\mu_3$ and $\mu_4$ are the third-- and fourth--order moments of
$\mW(x)$, while $\mu_2=\mu^2+\sigma^2$ is the analogous notation for the
second moment. In this expression one may regard the expectation values as
evaluated with $\mP(x,t)$ rather than with $\mQ(x,t)$, since the differences
are due solely to the change $\Gamma(\Wh(k))\to\Gamma(\Wh(k)+1)$ from
eq.~(\ref{eq:finteg}) to eq.~(\ref{eq:nuinteg}) and are subleading.  We can
now recognize a distintive mark of the RRT over the real line: for
sufficiently large time, the kurtosis of the average distribution profile
is certainly positive, since positive definite is the fourth moment of any
$\mW(x)$. Another important characteristic, which will be further
discussed later on, is the positivity of the ratio between the skewness
$\avg{\xi^3}_t$ and the third moment $\mu_3$ of $\mW(x)$.

The extension of the main results,
eqs.~(\ref{eq:avgdelta})--(\ref{eq:hcumul}), to the case of a generic
initial distribution are straightforward. In particular, to the cumulants
of $\mQ(x,t)$ one would have to add the cumulants of the initial
distribution, which are constant in time and therefore subleading.
Thus eqs.~(\ref{eq:avgdelta}) would get additive constants and 
eqs. (\ref{eq:hcumul}) would hold unchanged. This is the standard way to see
how the process forgets abouts the initial conditions (in a logarithmic
slow way).

\section{Profile fluctuations}

Let us assume that, for a given stochastic matrix $W$ and initial
distribution $Q(\ell,N)$, we can explicitly compute $P(\ell,t)$ and
$Q(\ell,t)$ at least for large $t$, as in the preceding section. To
compare the result to the length distribution in a single evolution
history, or very few of them, which is indeed our case, we need to gather
information also on the fluctuations of the profile of the length
distribution from one history to the other.

Typically, one would like to rely on the law of large numbers. For ergodic
Markov chains (with finitely many possible events) this law states that the
probability that the frequency of a certain event in a given history
differs from its equilibrium probability by any nonzero amount vanishes
when the history becomes infinitely long. In our case the elementary events
are the observed protein lengths and the frequency in a given history is
just the profile of the length distribution in a given evolution history.
The quantity $Q(\ell,t)$ discussed above is just the expectation value of
the profile, that is its average over all possible histories. In a
Markovian setup with finitely many possible events there would be no
difference between the profile of a specific history and its expectation
value in the $t\to\infty$ limit, which means vanishing profile fluctuations
in the limit and negligible ones for sufficiently large $t$. The stochastic
process at hand, however, is not Markovian, having the (very specific,
simple and itself random) RRT form of memory and has a number of possible
events in principle arbitrarily large. In such case we expect, thanks to
stronger forms of law of large numbers like the central limit theorem (and
have indeed verified in the example class of the preceding section), that
the average length distribution $Q(\ell,t)$ assumes, for $t$ large enough,
a universal nonconstant form which depends only on very general properties
of $W$.

What we need then is also that the fluctuations of the frequency for large
$t$ do not spoil completely the profile of its expectation value $Q(\ell,t)$.
Notice, for instance, that this is not true for random walks, not even when
they are recurrent (as generally true in one dimension, which is our case).
In other words, in the standard RW the frequency of times the walker visits
any given small region keeps fluctuating strongly from one very long
history to the other, never resembling the mean frequency profile. This is
due to the characteristic dispersion of order $\sqrt{t}$ of the RW, which
implies that each elementary event occurs an insufficient number of times of
order $1/\sqrt{t}$ to guarantee a good convergence of the frequency along a
single history (it would be even worse in $d>1$ dimensions).

On the contrary, the random memory of the RRT dramatically helps the
application of the law of large numbers, since the logarithmic time
dependence leads to much slower drift and diffusion, strongly reducing the
inpact of fluctuations on the length distribution.  One could say that the
length distribution is an almost ``self-averaging'' array of random
variables, which for sufficiently long time does not differ too much from
its expected value. Indeed, at least in the case of the abstract RRT, there
exist mathematically rigorous theorems about the convergence of the
chemical distance profile of any RRT towards a normal form \cite{maths}. In
this section we provide some quantitative numerical evidence of the same
property for lengths distribution profiles using a specific model
for $W(\ell|\ell')$.

We first revert to the realistic situation of lengths as positive integers
not smaller than some lower cutoff $\ell_{\rm min}\ge1$; next we consider
the following RRT process (written as computer pseudocode)
\begin{equation}\label{eq:generic}  \boxed{ 
  \begin{split}
    &\ell = {\sf integer~part~of} \,e^x\ell(n_t); \\
    &{\sf if} \;\ell \ge \ell_{\rm min}\;{\sf then}\; \ell(t+1) = \ell
  \end{split}}
\end{equation}
where $n_t$ is an integer chosen at random from $1$ to $t$ and $x$ is
extracted with the one--step pdf $\mW(x)$ over the continuous logspace;
finally we pick for $\mW(x)$ the maximum entropy form compatible with our
general setup, namely a Gaussian with mean $\mu$ and standard deviation
$\sigma$. This minimum bias choice could even be regarded as natural in
view of the many different ``microscopic'' and ``macroscopic'' mechanisms
on which the stochastic process should depend, as discussed in the
Introduction. However, we make it here mainly for numerical definiteness.
In any case the analysis of the preceding section and the discussion below,
at the end of this section, should make it clear that other choices of
$\mW(x)$ in the same class would lead to relative changes that vanish as
$1/\log t$, while preserving important characteristic properties like the
positivity of the kurtosis.

It is quite easy on modern personal computers to accurately simulate the
process (\ref{eq:generic}) by running many very long random histories. In
our simulations, we produced $10^5$ length distributions with the discrete
time $t$ running from $N_0\lesssim 50$ to $N=5\cdot 10^6$. For sake of
definiteness we started from $25$ initial lengths chosen at random from
$30$ to $50$ and set $\mu =0.16$ and $\sigma=0.19$.  This setup was
determined in such a way to fit the overall scales of the length
distribution in the SIMAP database, as will be discussed in the next
section. In particular, it turns out that the effects on the profiles of
the lower cutoff $\ell_{\rm min}$ are fully negligible, so that the
scale--invariant framework adopted in the preceding section should apply.
Indeed, one can also check that the discreteness of the lengths $\ell(t)$
does not play any significant role at all w.r.t. the continuous case.

For each distribution we computed the mean and
standard deviation in the variable $x=\log\ell$:
\begin{equation}\label{eq:extim1}
  \mu_t = \frac1{t}\sum_{j=1}^t x(j) \;,\quad
  \sigma_t^2= \frac1{t}\sum_{j=1}^t [x(j)-\mu_t]^2
\end{equation}
at prescribed intermediate values of $t$. Likewise, we computed the
skewness and kurtosis
\begin{equation}\label{eq:extim2}
  s_t = \frac1{t}\sum_{j=1}^t \xi^3(j) \;,\quad
  \kappa_t=  -3 + \frac1{t}\sum_{j=1}^t \xi^4(j)
\end{equation}
where as usual $\xi(j)=[x(j)-\mu_t]/\sigma_t$.

These four parameters are still random variables which fluctuate from one
RRT to the other. Moreover, except for the mean $\mu_t$, their average
values over all possible RRT realizations of $t$ steps do not coincide with
the corresponding parameters of the average distribution $\mQ(x,t)$, since
such average values receive contributions also from the profile
fluctuations.  Only the average $\avg{\mu_t}$ is given by the quantity
${\bar\mu}_t$ in the first eq.~(\ref{eq:avgdelta}). The differences between
$\avg{\sigma_t}$, $\avg{s_t}$, $\avg{\kappa_t}$ and ${\bar\sigma}_t$,
${\bar s}_t$, ${\bar\kappa}_t$ in the second eq.~(\ref{eq:avgdelta}) and
eqs.~(\ref{eq:hcumul}), respectively, cannot be even extimated with the help
of $\mQ(x,t)$ alone. This is true a fortiori for the fluctuations.
Therefore it is important to provide some (numerical) evidence on their
behaviour for large times. In particular, $s_t$ and $\kappa_t$, provide a
measure of the deviation from gaussianity of a given profile (we refer to
above--mentioned tmathematical literature for some rigorous bounds in the
case of abstract RRT's).

We also kept track of all the logspace profiles, after a suitable coarse
graining: we fixed beforehand a uniform binning grid of $K$ intervals of
width $h\ll1$ over a portion of the real line large enough to contain
almost all $\xi$ points produced ({\em e.g.} the interval $(-5,5)$ to
comprise all points within $5$ sigmas); then we computed the fraction $q_k$
of $\xi$ points in a given RRT that fall in the $k$-th interval of the grid.
At this stage using continuous or discrete lengths does make a difference,
since a binning grid too fine over the logarithms of integer lengths will
induce spurious fluctuations. Hence in the discrete case, for each integer
$j$ repeated $n_j$ times in a given length distribution we filled the real
interval $(j-1/2,j+1/2$ with $n_j$ double precision lengths chosen at
random; only after this smoothing we computed the distribution over the
regularly space grid in logspace. 

By construction, the average of the discretized density $q_k(t)/h$ over all
possible histories will reproduce the integral of the average profile
$\mQ(x,t)$ as a function of $\xi$ over the $k$-th interval of the grid.
Then an extimate of the profile fluctuations is the standard deviation of
$q_k(t)/h$ for each $k$.

With $q_k(t)$ we computed another important (and more robust) measure
of deviation from gaussianity, that is the entropy:
\begin{equation*}
  S_t = \log h + \sum_{k=1}^N q_k(t) \log q_k(t)   
\end{equation*}
In fact, in the $t\to\infty$ limit of an infinite RRT  and then
$h\to0$ of vanishing grid width, a Gaussian profile for $\xi$ would have
maximal entropy equal to $(\log2\pi +1)/2 =1.41893853\ldots$.

\begin{figure}[!htb]
  \vspace{-0.5truecm}\hspace{-0.4truecm}
  \begin{center}
    \includegraphics[width=0.50\textwidth]{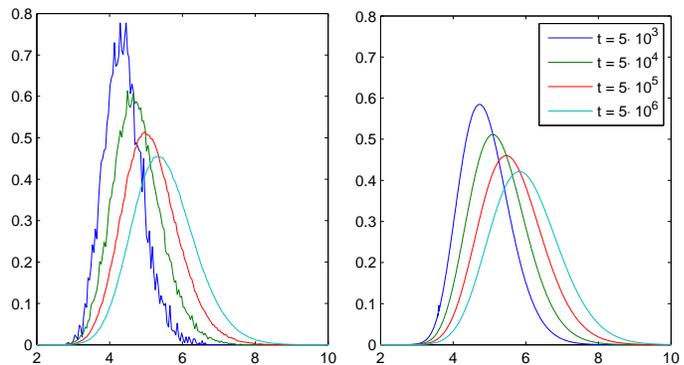}
    \caption{\label{fig:evol} Evolution of the logspace length distribution
      in a specific history (left panel) and on average (right panel),
      starting from the same initial conditions.}
  \end{center}
\end{figure}

In fig.~\ref{fig:evol} we show the evolution of the logarithmic length
distribution along a single history and, for comparison, the evolution of
the average profile $\mQ(x,t)$ obtained by numerically integrating through
FFT eq.~(\ref{eq:nuinteg}) and superposing the results as in 
eq.~(\ref{eq:Qsuperpose}).

\begin{figure}[!htb]
  \begin{center}
    \includegraphics[width=0.50\textwidth]{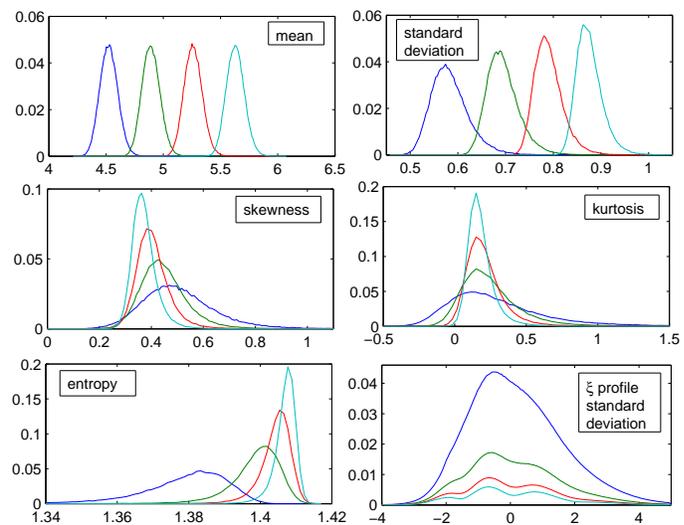}
    \caption{\label{fig:params} Distributions over $10^5$ sample histories
      of the indicated statistical extimators of the logspace length
      profile for the same times as in fig.~\ref{fig:evol}.}
  \end{center}
\end{figure}

In fig.~\ref{fig:params} we show the distributions of the statistical
extimators defined above for few values of $t$ equally spaced in logspace.
We see that the parameters that measure deviation from gaussianity, that is
$s_t$, $\kappa_t$ and $S_t$, have mean values that slowly tend to the
Gaussian values with smaller and smaller fluctuations as $t\to\infty$. The
convergence behaviour is roughly the ubiquitous one, $1/\log t$, with
variances that vanish faster than the peaks movement. Also the variance of
the standard deviation seems to slowly converge. On the other hand the
fluctuations of the mean do not appear to converge at all; this is
reflected in the reduction slower than $1/\log t$ of the standard deviation
of the $\xi$ profile fluctuations. In table~\ref{tab:deltas} we provide
further numerical evidence through the standard deviations over the $10^5$
sample histories of $\mu_t$, $\sigma_t$, $s_t$, $\kappa_t$ and $S_t$.

\begin{table}[!htb]
  \begin{center} {
      \begin{tabular}{|c||c|c|c|c|c|} \hline
        $t$ & $\Delta\mu_t$  &  $\Delta\sigma_t$ & 
         $\Delta s_t$ &  $\Delta\kappa_t$ & $\Delta S_t$ \\
         \hline
         $5\cdot 10^3$  & $\;0.0781\;$ & $\;0.0401\;$ & $\;0.1425\;$
         & $\;0.3840\;$ & $\;0.0131\;$ \\
         $\;5\cdot 10^4\;$ & $\;0.0784\;$ & $\;0.0343\;$ & $\;0.0927\;$
         & $\;0.2282\;$ & $\;0.0072\;$ \\
         $\;5\cdot 10^5\;$  & $\;0.0787\;$ & $\;0.0304\;$ & $\;0.0647\;$ 
         & $\;0.1440\;$ & $\;0.0045\;$ \\
         $\;5\cdot 10^6\;$  & $\;0.0785\;$ & $\;0.0277\;$ & $\;0.0482\;$ 
         & $\;0.0980\;$ & $\;0.0031\;$ \\\hline
      \end{tabular}}
    \caption{\label{tab:deltas}Standard deviations over $10^5$ samples of
      the statistical extimators of mean, standard deviation,
      skewness, kurtosis and entropy of the logspace length distribution.}
  \end{center}
\end{table}
These results remain qualitatively unchanged under generalization from the
Gaussian one--step pdf chosen above to a generic $\mW(x)$ of the class
discussed in the previous section. For fixed $\mu$ and $\sigma$ only
$t-$independent numerical variations appear due to the change in the higher
moments of $\mW(x)$. In particular the skewness can be made to assume
prevalently positive or negative values by choosing a $\mW(x)$ with
positive or negative third moment (while keeping the first moment $\mu>0$),
while the kurtosis distribution remain always peaked around positive
values, with a variance which appear to vanish faster than the mean. This is
in agreement with the properties of the average length distribution as
given by eq.~(\ref{eq:hcumul}).

In summary, very large RRT's over the space of possible protein lengths are
indeed almost auto-averaging objects and it is sensible to compare the
average properties of the random process to a few, or even a
single, realizations of it.

\section{Comparison with the observed length distributions}

To test our simple model we compare here the predicted length
distributions with the real length distributions of proteins observed in
nature. In this last decade the number of known protein sequences has been
rapidly growing and is still growing now at a steady pace. A huge number
of protein sequences coming from very many different species are now stored
in various databases.

In particular the {\bf SIMAP} \cite{simap} (Similarity Matrix of
Proteins, \e{http:$\slash\slash \text{mips.gfs.de}\slash
$genre$\slash$proj$\slash$simap}) database collects about all amino
acid sequences from public databases and completely sequenced
genomes. On September 2006 it was storing more than $5.5$ millions
of \e{not-redundant} proteins coming from more than $100000$ different
species.

\begin{table}[!htb]
  \begin{center} {\small
      \begin{tabular}{|lr|c|c|}
        
        \hline
        
        \parbox{0.1\textwidth}{\it kingdoms}  & & 
        \parbox{0.1\textwidth}{\it number of species} & 
        \parbox{0.1\textwidth}{\it number of proteins}   \\
        \hline
        
        \parbox{0.1\textwidth}{bacteria}  & \parbox{0.1\textwidth}{} & 
        \parbox{0.05\textwidth}{$11130$} 
        & \parbox{0.05\textwidth}{$2217301$}   \\
        \hline

	\parbox{0.1\textwidth}{viruses}  & \parbox{0.1\textwidth}{} & 
        \parbox{0.05\textwidth}{$14631$} 
        & \parbox{0.05\textwidth}{$319885$}   \\
        \hline

        \parbox{0.1\textwidth}{plants}  & \parbox{0.1\textwidth}{} & 
        \parbox{0.05\textwidth}{$31232$} 
        & \parbox{0.05\textwidth}{$1156929$}   \\
        \hline
        
        \parbox{0.1\textwidth}{animalia}  & \parbox{0.1\textwidth}{invertebrates} & 
        \parbox{0.05\textwidth}{$25951$} 
        & \parbox{0.05\textwidth}{$383760$}   \\
        
        \parbox{0.1\textwidth}{}  & \parbox{0.1\textwidth}{vertebrates} & 
        \parbox{0.05\textwidth}{$19341$} 
        & \parbox{0.05\textwidth}{$772605$}   \\
        \hline
        \parbox{0.1\textwidth}{environmental  samples}  & \parbox{0.1\textwidth}{} & 
        \parbox{0.05\textwidth}{$1453$} 
        & \parbox{0.05\textwidth}{$32591$}   \\
        \hline	 
   
        \parbox{0.1\textwidth} {synthetic}  & \parbox{0.1\textwidth}{} & 
        \parbox{0.05\textwidth}{$822$} 
        & \parbox{0.05\textwidth}{$14660$}   \\
        \hline	 
        
      \end{tabular} }
    \caption{\label{tab_bioclass} \small Number of species and
      proteins for each kingdom in SIMAP on September 2006.}
\end{center}
\end{table}

\begin{figure}[!htb]
  \hspace{-0.6truecm}
  \includegraphics[width=0.4\textwidth,angle=270]{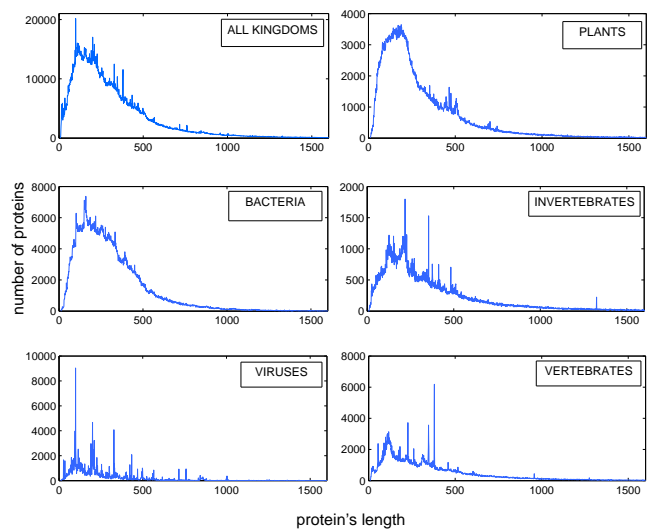}
  \caption{\label{fig:proteinbiospecies} {\small Length
      distributions of SIMAP proteins. Each box shows an enlargement
      of the (not normalized) length distributions of proteins
      coming from \e{all kingdoms} ($\avg{l} = 335$, $l_{max} =
      38031$), \e{bacteria} ($\avg{l} = 316.9$, $l_{max} = 36805$),
      \e{viruses} ($\avg{l} = 273.9$, $l_{max} = 7312$ ), \e{plants}
      ($\avg{l} = 314.5$, $l_{max} = 20925$), \e{invertebrates}
      ($\avg{l} = 416.1$, $l_{max} = 23015$), \e{vertebrates}
      ($\avg{l} = 397.1$, $l_{max} = 38031$).}}
\end{figure}

We report in Table~\ref{tab_bioclass} a coarse subdivision of all SIMAP
proteins and their corresponding species in five (non-standard) main
kingdoms: bacteria, viruses, plants, invertebrates (animalia) and
vertebrates (animalia). In fig. \ref{fig:proteinbiospecies} we provide
plots of the corresponding length distributions.

One can see that all SIMAP distribution profiles have a global similar
shape, with a well defined overall position and scale. There are however
also large fluctuations on smaller scales. In particular the curves of
viruses, invertebrates and vertebrates show very high and narrow peaks in
correspondence to certain specific values of length. Of course, on general
grounds, our model is too simple and generic to make predictions on other
than global properties of the profiles, so we should restrict to the lowest
moments or cumulants of the distribution and perform robust coarse graining
on the data for more refined analysis. We believe, in any case, that these
peaks are to a large extent spurious, being due to an over--representation
in the SIMAP database of those particular protein lengths. SIMAP in fact
contains a lot of proteins which not necessarly come from completely
sequenced genomes: this fact makes the collection of proteins not
homogeneous over the species present in the database and so it is possible
that certain peculiar lengths are more represented since they correspond to
proteins of many more different species than other lengths. If the
collection were homogeneous over the species, we would expect length
distributions without high narrow peaks and also less fluctuating in
general. At any rate, we verified that the global analysis reported below
is almost insensitive to the removal by hand of the high and narrow peaks.

The SIMAP database provides a very large sample of real proteins which
can be assumed to be statistically significant. We believe therefore
that it is sensible as a testbed for our model and we make the basic
assumption that the SIMAP length distributions for different kingdoms
as (almost) independent realizations of our stochastic process.  The
motivation is that different kingdoms have been going through almost
independent evolutional histories since a long time and, even if one
cannot forget that far enough in the past there was non distinction at
all, the main characteristic of the stochastic process of forgetting
the initial conditions suggests that at most a negligible trace remain
of the common remote past in each kingdoms distribution.

In table \ref{tab:protpars} we list the measured values of the mean,
standard deviation, skewness, kurtosis and entropy of the logarithmic
length distributions for the five kingdoms separately and for the comulative
all kingdoms distribution. Except for the entropy, these parameters can be
computed directly from the statistical extimators as in
eqs.~(\ref{eq:extim1}) and~(\ref{eq:extim2}) without any coarse graining.
To compute the entropy we performed a coarse graining in the logspace with
the same procedure of the preceding section. One can see that the kurtosis
is always positive, in accordance with the average property of the model
[eq.~(\ref{eq:hcumul})] and with the results of the previous section on the
fluctuations.  We also notice that the cumulative kurtosis is definitely
higher than the individual ones, due to the fluctuations in the lower
cumulants. Again, this is consistent with the interpretation of the kingdoms
distributions as independent realizations of the process. Except for the
viruses, the entropy is always close to the upper Gaussian limit, with the
plant distribution the closest to a normal form.

\begin{table}[htb]
  \begin{center} {
      \begin{tabular}{|c||c|c|c|c|c|} \hline
        {\it kingdom} & mean & ~s.d.~ & skewness  & kurtosis 
        & entropy \\\hline         
        bacteria      & $5.53$ & $0.68$ & $-0.20$ & $0.32$ & $1.408$ \\
        viruses       & $5.26$ & $0.79$ & $ 0.30$ & $0.53$ & $1.297$ \\  
        plants        & $5.44$ & $0.78$ & $-0.01$ & $0.04$ & $1.414$ \\
        invertebrates & $5.65$ & $0.87$ & $-0.03$ & $0.31$ & $1.409$ \\
        vertebrates   & $5.60$ & $0.89$ & $-0.18$ & $0.25$ & $1.394$ \\
        \hline 
        all kingdoms  & $5.49$ & $0.81$ & $-0.26$ & $0.63$ & $1.406$ \\ 
        \hline
      \end{tabular}}
    \caption{\label{tab:protpars} Global statistical indicators of the
      SIMAP length distributions in logspace.}
  \end{center}
\end{table}

\begin{figure}[!htb]
  \hspace{-0.6truecm}
  \includegraphics[width=0.4\textwidth,angle=270]{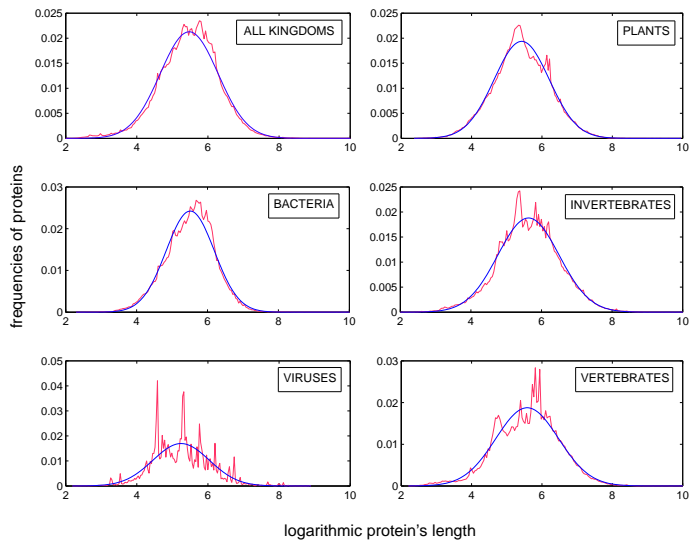}
  \caption{\label{fig:logspace} Length distributions of SIMAP proteins in
    the logarithmic lengths space. These (not normalized) distributions have
    been obtained through a uniform gridding with $200$ intervals over
    $x=\log\ell$. Before the change of variable from $\ell$ to $x$, we
    scattered the protein length values from integer to real to
    avoid the introduction of spurious fluctuations in the logarithmic
    space.  }
\end{figure}

In fig. \ref {fig:logspace} we show the Gaussian fits of the length
distributions. As expected from the data in table~\ref{tab:protpars}, these
fits appear rather good, apart from fluctuations, which are more important
when the entropy is lower, that is for viruses and vertebrates. Explaining
these fluctuations is beyond the scope of our model. Moreover, one must
remember that the SIMAP database is incomplete and, as discussed above,
probably biased toward particular species; these features contribute to
local irregularities.

With fine--tuned choices of the two main parameters of $\mW(x)$, $\mu$ and
$\sigma$, a specific large value of $t$ and values of the initial lenghts
in the range $30\,-\,50$, one can produce simulations with
eq.~(\ref{eq:generic}), like those reported in fig.~\ref{fig:params}, whose
distributions profiles fit well the peak positions and the sizes of the
SIMAP length distributions.

Our choice for the initial distribution is based on the quite natural
assumption that today's proteins are evolved from shorter peptide ancestors
\cite{protevol1, protevol2}. In any case, according to the model,
$t-$independent changes in the initial distribution might affects the final
distribution only by terms of relative order $1/\log t$.
 
We remark also that in our purely probabilistic framework no fine 
quantitative check can be performed, for some good reasons.

Firstly, even assuming that for each kingdom the proteins in the database
constitute a statistically significant fraction of the total existing in
nature, we do not know what the total number might actually be.  So we
cannot fix the precise value of the total discrete time of the stochastic
process. This does not lead to large uncertainities, though, since the
evolution is only logarithmic in this discrete time.

Secondly, the one--step pdf $\mW(x)$ governing the model can hardly have
any precise quantitative relation with the moltitude of microscopic and
macroscopic effects that drive the evolution. So one could not ascribe any
particular value to a specific functional form of $\mW(x)$ whose most
relevant free parameters were determined from data fitting. Rather we must
restrict to very general properties valid for a wide class of one--step
pdf's. This argument applies also to the lowest moments of $\mW(x)$, that
might very well differ in distinct independent processes.

Let us examine therefore in more detail to what extent the
model agrees with the observed distributions.

First of all, according to the model, the length distribution of a large
set of protein belonging to a {\em single} long evolutional history must be
almost a Gaussian in logspace, that is a lognormal over the lengths. As we
have just seen, this agrees quite well with the SIMAP distributions. The
approximate lognormality of protein sizes was observed several years ago on
much smaller data sets \cite{lognormds}.

Next there are the two scales of the fluctuations of the two parameters of
a Gaussian, namely the mean and the standard deviation, which were denoted
as $\mu_t$ and $\sigma_t$ in the previous section. Once a process
simulation is fine--tuned to produce the correct average values of $\mu_t$
and $\sigma_t$, their fluctuations have a scale which depends mildly on the
higher moments of $\mW(x)$, does not depend on $t$ in the case of $\mu_t$
and depend at most as $1/\log t$ in the case of $\sigma_t$.  These scales
agree fairly well with those observed over the SIMAP kingdom distributions
[see tables~\ref{tab:deltas} and ~\ref{tab:protpars}], at least when
$\mW(x)$ has skewness and kurtosis not too large. Notice that also the lower
cutoff $\ell_{\rm min}$ on the possible lengths acts as constraint on the
higher cumulants of $\mW(x)$. For instance, an excessively negative
skewness in $\mW(x)$ would typically lead to large left tails also in the
length distributions which are then abruptly cutoff at $\ell_{\rm min}$; such
abrupt cuts are absent in the observed data, as evident from
fig.~\ref{fig:logspace}.

Then there are the systematic deviations from gaussianity. The model
predicts a positive kurtosis for any $\mW(x)$ and the SIMAP data agree very
well with this.  Also the entropy is very close to the expected values,
except for the viruses, whose protein distribution is the least abundant,
has the smallest mean length and the largest relative fluctuations.
However, the data in table~\ref{tab:protpars} show also always negative
skewness, again except for the viruses. This characteristic cannot be
accounted for too easily in the model.

Indeed, as we have already noted, it is natural to assume that the average
protein length has been growing in time. This requires that the first
moment $\mu$ of the one--step pdf $\mW(x)$ is positive, that is that a
positive shifts in $x-$space prevails over negative ones. Now, from the
average relation~(\ref{eq:hcumul}) and the numerical analysis reported in
the preceding section, a prevalently negative skewness requires that the
third moment $\mu_3$ of $\mW(x)$ is negative. The simplest Gaussian
one--step pdf used to produce the results in fig.~\ref{fig:params} has by
definition $\mu$ and $\mu_3$ with the same sign, but other pdf's with
positive $\mu$ and negative $\mu_3$ are certainly possible. However, this
would mean even more negative skewness in $\mW(x)$, causing problems with
the lower cutoff $\ell_{\rm min}$, unless very large unrealistic values of
the initial lengths are assumed, which in turns would typically spoil the
overall scale fitting.  Thus, after all, there is tension on the model in
spite of its many free parameters.

Our simple model for the protein length distributions is based on RRT
embedded in the natural numbers, with the assumption of almost scale
invariance for the transition probability $W(\ell|\ell')$.  By this we mean
that the stochastic process is uniform in continuous logspace, with
translation invariance broken only by length discreteness and the lower
cutoff $\ell_{\rm min}$, in both cases with negligible effects. This is an
idealization suggested by simplicity (it translates the intuitive idea that
longer proteins can change more than shorter ones throughout evolutions)
and ease of analytic investigation of average properties. It has some
difficulties in accounting for negative skewness (viruses apart), but it
is in overall good agreement with observations, especially for the
positiveness of the kurtosis. This suggests to keep to a minimum the
modifications in more realistic models for $W(\ell|\ell')$. We describe one
minimal change in the next section.

\section{Intrinsic smooth cutoff on large lengths}

In the stochastic framework we have considered up to now, the vanishing of
the length distribution for large $\ell$ is determined by the slowly
drifting and diffusing character of the process with the assumption of
relatively small initial lengths.

On the other hand there are physical reasons to expect that very long
proteins are intrinsically less probable than shorter ones, in the sense
that the ``microscopic'' mechanisms that determine, upon countless
repetitions, the production of longer and longer proteins are eventually
limited by simple stability criteria: very long proteins, to be stable
against thermal fluctuations in the natural environment, must fold in the
biologically active form more ``tightly'' than shorter ones, as could be
measured by their growing spectral dimension \cite{davraf}; but this
requires more and more complex stereoscopic orderings while the building
blocks (amino acids at the lowest level and larger strutures at the second
and third level) are limited in number and typology. We could therefore
expect some form of smooth cutoff on long lengths, parametrized by a
stability scale $\ell_s$.
   
The minimal change on the model, as anticipated above, could therefore be
the following:
\begin{equation}\label{eq:smoothcut}
  W(\ell\,|\ell') = \ell^{-1} \, g(\ell/\ell_s)
  \,  \mW\big(\log(\ell/\ell')\big) 
\end{equation}
where $\mW(x)$ is the usual one--step pdf in logspace and $g(u)$ a smooth
function which is almost constant for $u\lesssim1$ and monotonically
decreases to zero for large $u$.

The random recursion corresponding to eq.~(\ref{eq:smoothcut}) is a
simple modification of eq.~(\ref{eq:generic})
 \begin{equation}\label{eq:smoothcut}  \boxed{ 
  \begin{split}
    &\ell = {\sf integer~part~of} \,e^x\ell(n_t); \\
    &{\sf if} \;\ell \ge \ell_{\rm min} \;{\sf and}\; 
    r < g(\ell/\ell_s) \;{\sf then}\; \ell(t+1) = \ell
  \end{split}}
\end{equation}
where, as before, $n_t$ is a random integer from $1$ to $t$ and $x$ is
extracted from $\mW(x)$, while the new random number $r$ is extracted
uniformely in the interval $(0,g_{\rm max})$, with $g_{\rm max}=g(\ell_{\rm
  min}/\ell_s)$ the assumed largest value of the function $g$.

Another possibility could be to introduce an explicit $\ell-$dependence in
$\mW(x)$, in such a way that length reductions ($x<0$) become more
probabale than length growths ($x>0$) for large enough lengths.  In this
case the random recursion would be the same of eq..~(\ref{eq:generic}); the
only change is that $x$ is extracted in a weakly non scale--invariant way
from a one--step pdf $\mW(x;\ell)$.
   
Once some specific form for $\mW(x)$ and $g(u)$, or for $\mW(x;\ell)$, is
chosen, simulations with weakly broken scale invariance can be performed as
easily as before. Since there are now more tunable parameters, it is almost
obvious that data fitting can be improved. From a purely quantitative point
of view this better fits have little significance.  Instead we want to
stress the main new qualitative aspects: the smooth cutoff typically 
induces shorter right tails in the simulated distribution, thus slightly
reducing both skewness and kurtosis. If the one--step pdf has the right
characteristics, it is possible to obtain almost always length
distributions with still positive kurtosis but negative skewness after a
few million steps. The cutoff prevents the formation of proteins too long,
thus allowing to reproduce the observed mean length and length variance
Typically $\ell_s$, which by construction provides the scale of the
rightmost tail, need to be choosen between $5000$ and $10000$, depending on
other details of the model. 

It is also interesting to observe that the positive skewness of length
distribution of the viruses does not constitue a problem to the above
scenario, since the overall size of this distribution is smaller than the
others and might very well be too small to feel the effects of the smooth
cutoff on higher lengths.

\section{Conclusions and outlook}

In this work we have described a simple stochastic framework for the
theoretical modeling of the evolution of protein lengths. It is based on
the idea of Recursive Random Trees over the set of natural numbers. RRT's
represent the simplest formal implementation of the main feature of the
evolution process: new biological material is produced through
modifications of the biological material already existing. In the case of
proteins the full space over which the RRT grows is that of all amino acid
sequences, but it can be reduced to more tractable spaces when only
specific observables are considered, as is the case of the protein lengths.

Of course, the details of the stochastic process, as encoded in the
conditional probabilities, are practically out of reach, due to the
moltitudes of natural causes ranging from biochemical interactions to
selection mechanisms in varying environments. The relevance of the
stochastic framework is based therefore on the concept of universality;
namely that, under the law of large numbers, statistical coarse grained
observations tend to take universal forms which depend only on few
fundamental features of the stochastic process. In the case at hand, the
main features are the auto-averaging property of RRT's and the approximate
scale invariance of the one--step transition probability; they imply the
universal properties that protein length distributions are almost lognormal,
with positive kurtosis and a specific scale for the overall deviations
from exact gaussianity.

There are several routes for improvements. First of all, the choice of
RRT's (which have a uniform probability over all nodes of the tree for the
attachment of the new node) is by itself an ideal simplification. In a more
realistic setup one should consider differently weighted nodes in order to
mimic certain aspects of the evolution process such as selection and
differentiation. Then there are many more observables other than
the distribution length in protein databases such as SIMAP. The global
statistical analysis of the SIMAP protein homology network carried through
in ref.\cite{miccio} shows several interesting features which deserve to be
studied within some generalization of the stochastic process described here.

\section{Acknowledgments}
We are thankful to Thomas Rattei for his kind permission to access the
SIMAP database.

\end{document}